\documentclass{acm_proc_article-sp}
\usepackage{verbatim}

\newcommand\T{\rule{0pt}{2.6ex}}
\newcommand\B{\rule[-1.2ex]{0pt}{0pt}}

\begin{document}

\title{Analysing Complexity of XML Schemas \\ in Geospatial Web Services}
\numberofauthors{3}

\author{
\alignauthor
Alain Tamayo\\
       \affaddr{Institute of New Imaging Technologies}\\
       \affaddr{Universitat Jaume I, Spain}\\
       \affaddr{Ave Vicent Sos Baynat, SN, 12071, Castell\'on de la Plana}\\
       \email{atamayo@uji.es}
\alignauthor
Carlos Granell\\
      \affaddr{Institute of New Imaging Technologies}\\
       \affaddr{Universitat Jaume I, Spain}\\
       \affaddr{Ave Vicent Sos Baynat, SN, 12071, Castell\'on de la Plana}\\
       \email{carlos.granell@uji.es}
\alignauthor 
Joaqu\'in Huerta\\
      \affaddr{Institute of New Imaging Technologies}\\
       \affaddr{Universitat Jaume I, Spain}\\
       \affaddr{Ave Vicent Sos Baynat, SN, 12071, Castell\'on de la Plana}\\
       \email{huerta@uji.es}
}

\maketitle

\begin{abstract}
XML Schema is the language used to define the structure of messages exchanged between OGC-based web service clients and providers. The size of these schemas has been growing with time, reaching a state that makes its understanding and effective application a hard task. A first step to cope with this situation is to provide different ways to measure the complexity of the schemas. In this regard, we present in this paper an analysis of the complexity of XML schemas in OGC web services. We use a group of metrics found in the literature and introduce new metrics to measure size and/or complexity of these schemas. The use of adequate metrics allows us to quantify the complexity, quality and other properties of the schemas, which can be very useful in different scenarios.
\end{abstract}




\section{Introduction}
Service-Oriented Architecture (SOA) is widely used in the Geographic Information Systems (GIS) field to access geospatial data. This architecture presents an approach for building distributed systems that deliver application functionality as services to either end-user applications or other services. One of the main requirements when building systems based on SOA is \textit{interoperability}, which ensures that information can be exchanged in a way that can be understood by service providers and consumers. 

Interoperability in the GIS field is achieved by using standards \cite{book:goodchild}. Some widely known standards are: Web Map Service Implementation Specification (WMS) \cite{ogc:wms}, Web Coverage Service Interface Implementation Specification (WCS) \cite{ogc:wcs2} and Web Feature Service Implementation Specification (WFS) \cite{ogc:wfs2}, among others. These specifications are known as a whole as OGC\footnote{Open Geospatial Consortium}  Web Services (OWS), and they allow GIS clients to access geospatial data without knowing details about how this data is gathered or stored. These specifications define the interface of the operations that must be supported by a web service provider and the structure of messages exchanged between providers and web service clients using XML Schema \cite{w3c:schemas1}\cite{w3c:schemas2}. Nowadays, when data is increasingly available at growing rates, services play a critical role as entry points to access and manage this data.

The efficient processing of XML messages has a great influence in the overall performance of concrete implementations of these services. The implementation of XML processing for the OWS schemas is a complicated task because the size of the schemas has been growing with time, reaching a state that makes developers' work very difficult. Writing code to manipulate the resulting XML instances is complex whether we write this code manually or use a code generator. The first option is recognized to be difficult and error-prone producing code that is hard to modify and maintain \cite{book:mclaughlin}. The second option based on the use of code generators usually produces a large number of classes that either offers a poor performance or occupy a large disk space that limits its applicability in certain environments with constrained resources, such as mobile devices. 

An example of the effect of complexity in a real implementation is the OX-Framework \cite{proc:broring} , which defines an architecture to access data located in several kinds of OWS servers. The framework includes a client whose binary distribution occupies 58.8 MB, of which 46.4 MB is binary code for XML processing. This code, generated with XMLBeans\footnote{http://xmlbeans.apache.org/}, represents 79\% of the size of the distribution and contains 33,437 classes. The framework libraries present a lot of redundancy on the generated XML processing code, as common dependencies of service specifications have not been properly factorised in the final code. In any case, even when eliminating this redundancy will largely decrease the overall size of the framework, the fraction of the code related to XML processing will still be the largest part. 

In order to cope with these problems a first step is to measure the complexity of the schemas. For this reason, in this paper we present an analysis of the complexity of OWS schemas using a group of metrics found in the literature. We also introduce three new metrics to measure the complexity introduced by the use of the XML Schemas subtyping mechanisms. The use of adequate metrics allows us to quantify the complexity, quality and other properties of the schemas. The remainder of the paper is structured as follows. Next section presents a brief introduction to XML Schemas. Section 3 introduces OGC Web Services. After this, Section 4 presents the complexity analysis. In this section, we present metrics and their values for the considered specifications. Section 5 presents related work. Last, we present conclusions of our work.

\section{XML Schema}

XML Schema files are used to assess the validity of well-formed element and attribute information items contained in XML instance files\cite{w3c:schemas1}\cite{w3c:schemas2}. An XML Schema document mainly contains components in the form of complex and simple type definitions, element declarations, attribute declarations, group definitions, and attribute group definitions.

This language allows users to define their own types, in addition to a set of predefined types defined by the language, in the form of complex and simple types. Elements are used to define the types content and when global, to define which of them are valid as top-level element of a XML instance document. Figure  \ref{fig1} shows a fragment of an XML Schema file, which contains the declaration of three global complex types and a global element. Figure \ref{fig1} also shows how recursive structures can be defined as for instance \textit{ContainerType} that contains an element of the same type.

\begin{figure}[!h]
\raggedright
<complexType name=``BaseType''>\\
\hspace{3mm}<sequence>\\
\hspace{6mm}<element name=``baseElement''\\
\hspace{9mm} type=``string''minOccurs=``1''/>\\
\hspace{3mm}</sequence>\\
\hspace{3mm}<attribute type=``string'' use=``required''\\
\hspace{6mm} name=``id''/>\\
</complexType>\\ 
\vspace{5mm}
<complexType name=``ChildType''>\\
\hspace{3mm}<complexContent>\\
\hspace{6mm}<extension base=``BaseType''>\\
\hspace{9mm}<sequence>\\
\hspace{12mm}<element name=``childElement'' \\
\hspace{15mm} type=``string''/>\\
\hspace{9mm}</sequence>\\
\hspace{6mm}</extension>\\
\hspace{3mm}</complexContent>\\
</complexType>\\ 
\vspace{5mm}
<complexType name=``ContainerType''\\
\hspace{3mm}<sequence>\\
\hspace{6mm}<element name=``containerElement'' type=``BaseType''\\
\hspace{9mm}maxOcurrs=``unbounded''/>\\
\hspace{6mm}<element name=``recursiveElement'' \\
\hspace{9mm}type=``ContainerType'' minOcurrs=``0''/>\\
\hspace{3mm}</sequence>\\
</complexType>\\ 
\vspace{5mm}
<xs:element name=``container'' type=``ContainerType'' /> \\ 
\caption{XML Schema file fragment.}\label{fig1}
\end{figure}

XML Schema provides a derivation mechanism to express subtyping relationships. This mechanism allows types to be defined as subtypes of existing types, either by extending the base types content model in the case of derivation by extension (\textit{ChildType} in Figure \ref{fig1}); or by restricting it, in the case of derivation by restriction. Apart from type derivation, a second subtyping mechanism is provided through substitution groups. This feature allows global elements to be substituted by other elements in instance files. A global element E, referred to as \textit{head element}, can be substituted by any other global element that is defined to belong to the E's substitution group.

Schema components defined in a schema document can be reutilized in other documents through the use of \textit{include} and \textit{import} tags. Components defined in the same namespace can be accessed in a schema file by using the include tag, which specifies in the \textit{schemaLocation} attribute where the external schema is located. Similarly, components defined in a different namespace may be accessed by \textit{importing} the namespace and optionally specifying where the external schema is located.

\section{OGC Web Services}

As mentioned before, geospatial web services interfaces, defined by OGC, describe both data formats and structure of messages exchanged by web services clients and providers using XML Schema. Table  \ref{tab1} shows a short description of the web service interfaces used later on this paper. 

For each service specification we have chosen the last approved version at the time of writing this paper. For some of them, such as those related with sensors, new versions will be available soon, but their schemas are not still available on the official schemas repository in the OGC website\footnote{http://schemas.opengis.net/}. These specifications are included in the Sensor Web Enablement (SWE) initiative, a framework of open standards for exploiting Web-connected sensors and sensor systems of all types \cite{ogc:swe}. A comprehensive review of the new features included on SWE can be found in  \cite{article:broring}.

\begin{table}[!h]
\centering
\caption{Geospatial web service interfaces}\label{tab1}
\begin{tabular}{|p{2.7cm}|p{4.8cm}|} \hline
Name \T \B &Description\\ \hline
Web Map Service (WMS)& It produces maps of spatially referenced data dynamically from geographic information \cite{ogc:wms}\\ \hline
Web Feature Service (WFS) & It allows a client to retrieve and update geospatial data encoded in GML format \cite{ogc:wfs2}\\ \hline
Web Coverage Service (WCS)& It provides access to rich sets of spatial information, in forms useful for client-side rendering, multi-valued coverages, and input into scientific models \cite{ogc:wcs2}\\ \hline
Sensor Observation Service (SOS)& It provides an API to retrieve sensor and observation data \cite{ogc:sos}\\  \hline
Web Processing Service (WPS)&It defines a standardized interface to publish geospatial processes \cite{ogc:wps} \\ \hline
Sensor Planning Service (SPS)& It defines interfaces for queries that provide information about the capabilities of a sensor and how to task the sensor\cite{ogc:sps} \\\hline\end{tabular}
\end{table}

\begin{figure}[!h]
  \centering
  \includegraphics[scale=0.59]{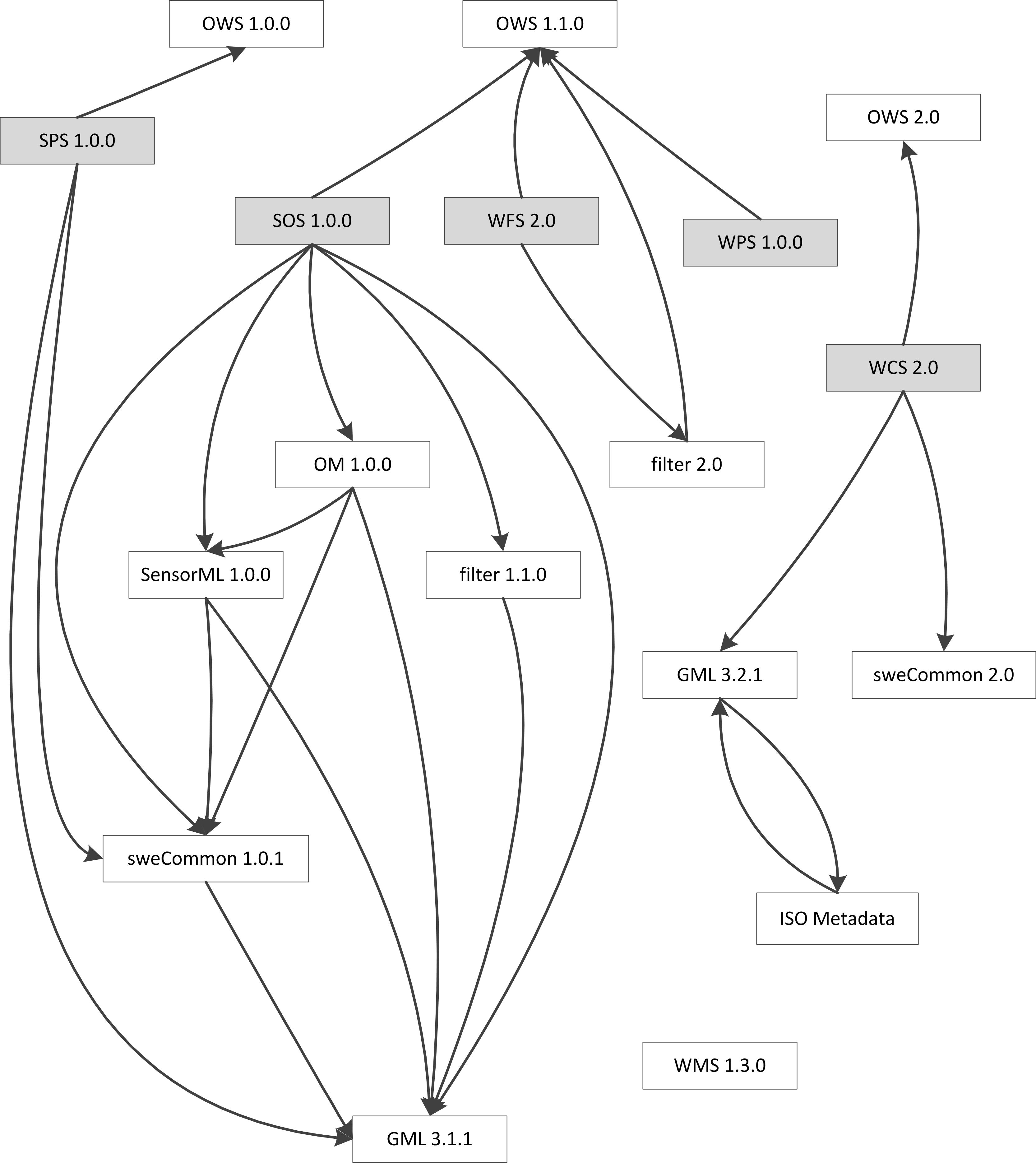}\\
  \caption{Dependencies between OGC specifications}\label{fig2}
\end{figure}

An important point about service specifications is that, as reusability is an important requirement when building software systems, they have been built on the foundation provided by other standard specifications and data models such as the  Geography Markup Language (GML)\cite{ogc:gml} and OWS Common \cite{ogc:common} specifications (Figure  \ref{fig2}). The reutilization of existing components simplifies the specification design task, but it also brings the complexity of the reutilized specifications into the other specifications as well. The schemas in OWS services are organized in a folder structure containing at the same level a folder for every specification. Each of them contains a folder for every version of the specification. Last, this folder contains the schema files, which we call \textit{main specification schemas}.  We differentiate these schemas from \textit{external schemas}, which are included or imported from the main specification schemas.

\section{Complexity Analysis}

In this section we present the complexity analysis for the schemas in the specifications listed in Table  \ref{tab1}. 

\subsection{Metrics}

For our study we have selected a set of metrics taken mainly from \cite{proc:lammel} and \cite{proc:mcdowell}. According to \cite{proc:lammel} metrics are categorized in \textit{XML-agnostic}, \textit{XSD-agnostic} and \textit{XSD-aware}. \textit{XML-agnostic metrics} do not consider any XML-related information. In this category we include for our analysis:
\begin{itemize}
\item \textit{Lines of Code (LOC)}: Total number of lines of code on the specifications' schemas. 
\item \textit{	Number of files (\#F)}: Total number of files related to the specification. Here, we consider recursively all of the files referenced through include and import XML schema statements. 
\end{itemize}
\textit{XSD-aware metrics}, consider metrics concerned with schema information. Here we consider a larger number of metrics:
\begin{itemize}
\item \textit{Number of complex types (\#CT)}: All complex types, including global and anonymous complex types.
\item	\textit{Number of simple types (\#ST)}: All simple types, including global and anonymous simple types
\item \textit{Number of global elements (\#EL)}: All global element declarations.
\item	\textit{Number of global model groups (\#MG)}: All global model groups definitions
\item	\textit{Number of global attributes (\#AT)}: All global attribute declarations
\item	\textit{Number of global attribute groups (\#AG)}: All global attribute groups definitions
\item	\textit{Number of global items (\#ALL)}: Number of global schema components (types, elements, model groups, attributes and attribute groups). 
\item	\textit{Wildcards}: Number of times wildcards are used.
\item	\textit{C(XSD)}: This metric calculates a complexity weight taking into account the internal schema components' structure \cite{article:basci}.
\item	\textit{Use of subtyping features of XML schema}:   Here, we consider first basic counting metrics such as the number of times certain features such as substitution groups, specialization by restrictions and specialization by extension, are used. After this, we introduce three new metrics to measure the influence of subtyping on the complexity of the schemas: \textit{Data Polymorphism Rate} (DPR), \textit{Data Polymorphism Factor}(DPF) and \textit{Schemas Reachability Rate} (SRR).
\end{itemize}
\textit{XSD-agnostic metrics} do not consider any information related with XML schema but use XML-related information. Examples of these metrics are \textit{Number of XML nodes} or \textit{Number of XML nodes for annotations} \cite{proc:lammel}. In our analysis we do not include metrics in this category because they are a measure of the size of the schemas, and this property is already measured by other considered metrics (e.g. LOC, \#F).

To calculate the metric values for a given specification, all of the schema files imported directly or indirectly from the main specification schemas are included. For this reason, it must be noticed that an actual implementation may not provide support for all of the schema components included in all of the schema files, but only for a subset of it. The size of this subset will depend directly from the specific implementation requirements.

There are a lot of other metrics that could be included in this study, but we chose those we consider more relevant. In some cases we discard some metrics because the information they provide is similar to that provided by other metrics. Next, we present in detail the \textit{C(XSD)} metric because it is more sophisticated than the other metrics considered in this study. The new metrics introduced in this paper will be explained in Section 5.4.

\subsection{C(XSD) Metric Definition}

In our study we include the metric presented in \cite{article:basci} that measures the schemas' complexity based in its internal structure, opposed to the metrics presented so far that limit themselves to just count schema components or features. It pays special attention to the use of recursive structures as a source of complexity to schema users. A complexity value, or \textit{weight}, is calculated for each schema component as an aggregation of the weights of the components it contains. The overall value of the metric is calculated with the following formula:
\begin{equation} \label{eq1}
\begin{split}
C(XSD) = \sum_{i=1}^{N}C(E_{gi}) + \sum_{j=1}^{M}C(A_{gj}) +\sum_{t=1}^{K}C(EG_{gt})  \\ 
               + \sum_{r=1}^{P}C(AG_{gr}) + \sum_{s=1}^{R}C(CT_{gs}) + \sum_{q=1}^{Q}C(ST_{gq})
\end{split}
\end{equation}

where, the first two terms are the summation of weights of global element and attribute definitions respectively. The remaining terms are summation of weights of global unreferenced model groups, attribute groups, complex and simple types that are declared/defined in the main specification schemas. The values N, M, K, P, R and Q are the number of global elements, attributes, unreferenced element groups, unreferenced attributes groups, unreferenced complex types and unreferenced simple types respectively.  In the second group of terms only unreferenced components are considered to avoid counting them several times as they are used in the declaration of global elements and attributes.

In \cite{article:basci} formulae are provided to calculate the weight of the different types of schema components. For example to calculate the weight of a complex type we use the following formula:

\begin{equation}
\begin{split}
w_{type} = w_{baseType} \pm{} [\sum_{i=1}^{N}C(E_{gi}) + \sum_{j=1}^{M}C(A_{gj}) \\ +\sum_{t=1}^{K}C(EG_{gt}) + \sum_{r=1}^{P}C(AG_{gr})] + NRC*R
\end{split}
\end{equation}
where, $w_{baseType}$ \footnote{The notation $w_{typename}$ is equivalent to $C(CT_{typename}) $}  is the weight of the base type. If derivation is not explicit (\textit{anyType} is the base type) the weight of the base type is 1. If we are in the case of derivation by extension,\textit{ N, M, K, P,} are the number of not inherited local or referenced elements, attributes, element and attribute groups referenced, that are not related to any element containing recursion. The sum of all these values is added to the weight of the complex type. If the complex type is derived by restriction, \textit{N, M, K, P}, are the corresponding number of schema components not inherited from the base type, and its weight is subtracted from the weight of the base type. In both cases \textit{NRC} is the number of child elements that contains recursion; and \textit{R} is an integer value greater or equal than 1 that can be understood as the weight given to recursion in a schema set. 

\section{Results}

The results of applying the metrics mentioned before to the specification schemas listed in Table \ref{tab1} are shown in the following subsections. 

\subsection{XML-Agnostic Metrics}

We start with \textit{XML-agnostic metrics}, which are those that do not consider XML-related information. The total amount of lines of code (LOC) and the number of files (\#F) give us a raw idea of the size of a given schema set. Table \ref{tab2} shows the value of the metrics for the considered OWS specifications. According to the categorization for LOC values presented in \cite{proc:lammel}, a schema set with between 10,000 and 100,000 LOC is considered \textit{large}. Values between 1,000 and 10,000 correspond to \textit{medium}-sized schemas. And values between 1,000 and 100 correspond to \textit{small} schemas. There are also categories for \textit{mini} schemas, below 100 LOC, and \textit{huge} schemas, above 100,000 LOC. Considering the overall values, 3 out of 6 of the specifications are considered large, two of them are considered medium-sized and the last one is considered small.

\begin{table}[!h]
 \begin{center}
 \caption{Lines of code (LOC) and number of files (\#F)} \label{tab2}
   \begin{tabular}{| p{0.8cm} | p{0.8cm} | p{0.8cm} |p{0.8cm} |p{0.8cm}|p{0.8cm} |p{0.7cm} | }
     \hline
  \T \B   & SOS 1.0.0 & WFS 2.0 & WCS 2.0 & SPS 1.0.0 & WPS 1.0.0 & WMS 1.3.0 \\ \hline
     $LOC$ & 17,581 & 3,631 & 15,416 & 14,361 & 3,326 & 761 \\ \hline
     $\#F $& 87 & 23 & 87 & 73 & 29 & 3 \\ \hline
   \end{tabular}
\end{center}
 \end{table}
WCS and the specifications related with sensors, SOS and SPS, exhibit the higher values for the metrics. It is not a coincidence that they are the ones with higher number of dependencies from other specifications (Figure \ref{fig2}). On the other hand, WMS turns out to be the simplest (which is maybe why is the most widespread) of the specifications being, in terms of lines of code, about 20 times smaller than SOS. WMS does not depend on any major external schema for its definition. From the results might be a little surprising such low figures for WFS. This is because WFS schemas are not linked explicitly to GML schemas, in which case the values for LOC and \#F would be very similar to those of WCS. 

\subsection{XSD-Aware Simple Metrics}

\begin{figure}[!h]
  \centering
  \includegraphics[scale=0.13]{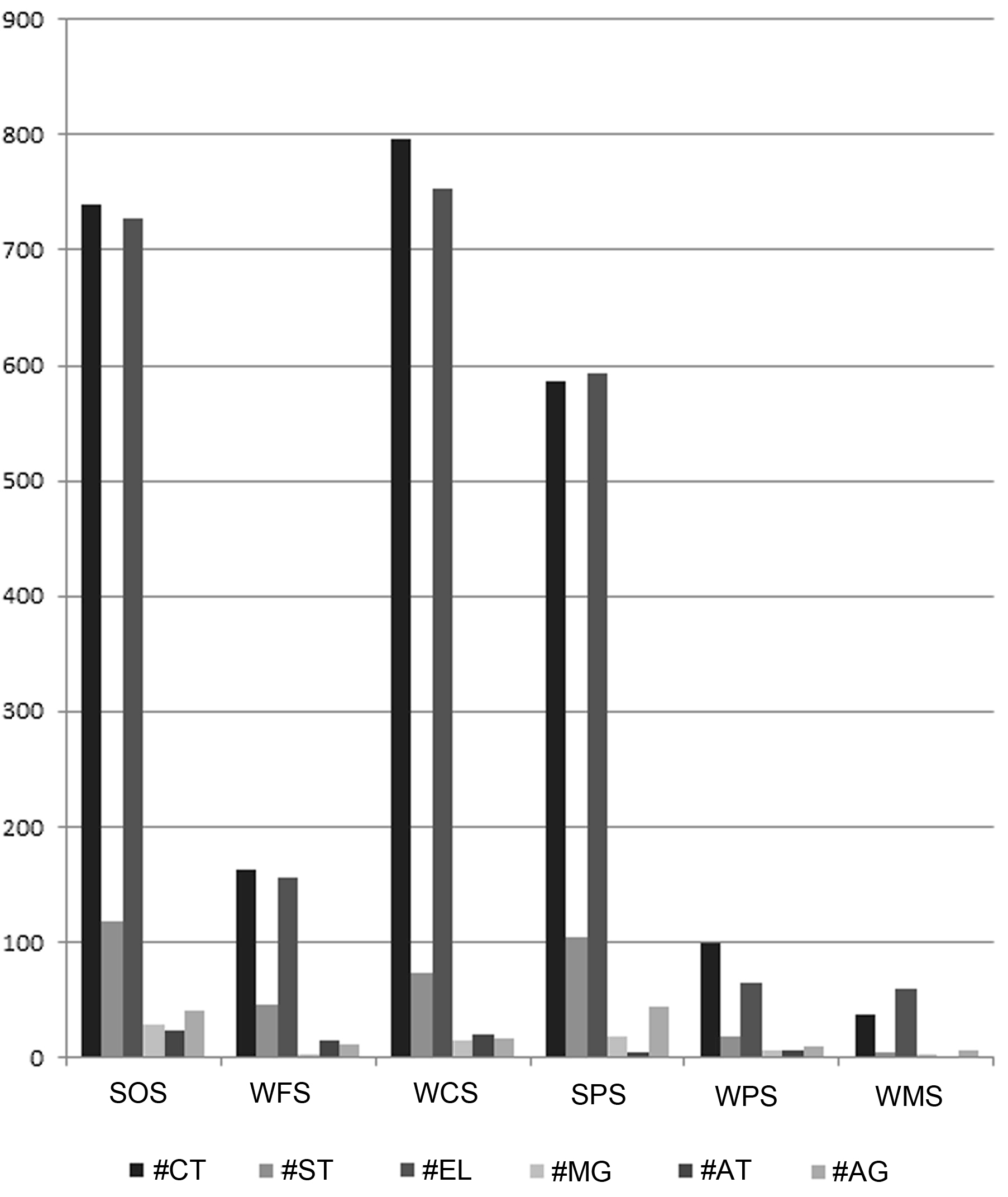}\\
  \caption{Number of main XML Schema components}\label{fig4}
\end{figure}

In this section we present the results of applying the XSD-aware metrics that count the number of main schema components. We start with the number of complex types (\#CT) which is considered paramount for measuring complexity because it measures the number of structured concepts modelled by the schemas \cite{proc:lammel}. Also, types are the fundamental concept when schemas are used to write (or generate) XML data binding code. The \#CT metric includes global complex types, as well as anonymous complex types. The number of complex types by specification is shown in Table \ref{tab3} and compared with other metrics in Figure \ref{fig4}.

\begin{table}[!h]
 \begin{center}
 \caption{Number of Complex Types (\#CT)} \label{tab3}
   \begin{tabular}{| p{0.8cm} | p{0.8cm} | p{0.8cm} |p{0.8cm} |p{0.8cm}|p{0.8cm} |p{0.7cm} | }
     \hline
 \T \B    & SOS 1.0.0 & WFS 2.0 & WCS 2.0 & SPS 1.0.0 & WPS 1.0.0 & WMS 1.3.0 \\ \hline
     $\#CT $ & 740 & 163 & 797 & 587 & 99 & 38 \\ \hline
   \end{tabular}
\end{center}
 \end{table}

Schemas with \#CT in the range 256-1,000 are considered \textit{large}, in the range 100-256 are considered \textit{medium} and \textit{small} with \#CT between 32 and 100 \cite{proc:lammel}.  The values of the overall metrics for all of the 6 specifications belong to these three ranges, 3 of them are large, one of them is medium-sized and the rest are small schemas. Again, WCS, SOS and SPS are among the most complex schemas and WMS is the simplest. As complex types model concepts, we can state that higher values of the metric imply higher conceptual complexity. 

A categorization of the schemas based on the number of other schema components is not provided in the literature. Nevertheless, they can give us some idea of schemas size and how often these features are used in the specifications.

\begin{table}[!h]
 \begin{center}
 \caption{Main XML features metrics (except \#CT)} \label{tab4}
    \begin{tabular}{| p{0.8cm} | p{0.8cm} | p{0.8cm} |p{0.8cm} |p{0.8cm}|p{0.8cm} |p{0.7cm} | }
     \hline
  \T \B  & SOS 1.0.0 & WFS 2.0 & WCS 2.0 & SPS 1.0.0 & WPS 1.0.0 & WMS 1.3.0 \\ \hline
     \#ST     & 118  & 46   & 74   & 105  & 18  & 5 \\ \hline
     \#EL     & 727  & 156  & 754  & 593  & 64  & 60 \\ \hline
     \#MG     & 28   & 3   & 14   & 19   & 7   & 2 \\ \hline
     \#AT     & 23   & 15    & 20    & 5    & 7   & 0 \\ \hline
     \#AG     & 40   & 12   & 17   & 44   & 10  & 7 \\ \hline
     \#ALL & 1498 & 350 & 1625 & 1266 & 179 & 80 \\ \hline
   \end{tabular}
\end{center}
 
 \end{table}

Table \ref{tab4} shows the overall values of the metrics for these schema components, which are also included in Figure \ref{fig4}. These values reinforce the idea of having a clear differentiation between a first group containing large specifications (SOS, SPS and WCS), a second group of medium-sized specifications (WFS), and a third group containing small specifications (WPS and WMS). In Figure \ref{fig4} we can observe the correlation that exists between the values of the metrics. This observation suggests that the coding style used in the schemas is consistent through all of the specifications. 

We would like to point out again that an actual implementation of WFS would require the use of other schemas, making them ascend to the category of large schemas.

Last, we count wildcards, which allow schema designers to specify extensibility points using \textit{<any>} and \textit{<anyAttribute>} tags. Using these tags in a complex type definition indicate that any global element or attribute can occupy that place in an instance document. The scope of the valid elements the wildcard is substituted for can be constrained by their namespace. 

The use of wildcards is widespread with the purpose of keeping schemas extensible, but they greatly complicate the processing of instance files.
A discussion about why the use of wildcards should be avoided when designing web service interfaces can be found in \cite{article:pasley}. We just want to highlight the fact that when parsing an XML instance we cannot be sure of what we will find in the place of the wildcards, so we must be ready to find almost anything. This obviously makes the source code for processing the instances files more complicated. If instead of writing the code manually we use a code generator, the presence of wildcards limits their possibilities of performing optimizations.

\begin{table}[!h]
 \begin{center}
 \caption{Use of wildcards} \label{tab5}
   \begin{tabular}{| p{1.3cm} | p{0.7cm} | p{0.7cm} |p{0.7cm} |p{0.7cm}|p{0.7cm} |p{0.7cm} | }
     \hline
  \T \B    & SOS 1.0.0 & WFS 2.0 & WCS 2.0 & SPS 1.0.0 & WPS 1.0.0 & WMS 1.3.0 \\ \hline
     Wildcards      & 13   & 13    & 5    & 9    & 0   & 0 \\ \hline
   \end{tabular}
\end{center}

 \end{table}
 
 Table \ref{tab5} shows how wildcards are used in the specifications. In this case, only the specifications already labelled as large make use of this feature.

\subsection{C(XSD)}

C(XSD) was introduced in Section 4.1. The metric calculates a weight for each schema component taking the internal structure of the component into account. Table \ref{tab6} shows the value of the metric. 

\begin{table}[!h]
 \begin{center}
 \caption{C(XSD) values for OWS specifications} \label{tab6}
   \begin{tabular}{| p{0.8cm} | p{1.0cm}| p{0.8cm} | p{1.0cm} |p{0.8cm} |p{0.6cm}|p{0.7cm} | }
     \hline
     \T \B  & SOS 1.0.0 & WFS 2.0 & WCS 2.0 & SPS 1.0.0 & WPS 1.0.0 & WMS 1.3.0 \\ \hline
     $C_{XSD}$   & 261,238 + 2,381R  & 1,960 + 16R & 209,997 + 1,171R & 96,451 + 885R    & 1,578 + 2R &  707 + 3R  \\ \hline
   \end{tabular}
\end{center}
 \end{table}

Previous metrics have shown that SOS was among the most complex specifications. C(XSD) shows that considering the internal structure of the schema components SOS is significatively more complex than the next specification, WCS. This is motivated by the fact that SOS has the higher number of dependencies from other specifications (GML, SWE  Common, OWS Common, O\&M\cite{ogc:om}, SensorML\cite{ogc:sml}, etc.), which are complex specifications as well. SensorML and O\&M contain the most complex schema components if analysed individually. The schema component with the higher individual value for C(XSD) is \textit{Component} in SensorML with 23,016 + 219R. Coincidentally, this element contains the highest number of recursive branches in its definition.

\subsection{Subtyping Mechanisms}

XML Schema subtyping mechanisms were introduced in Section 2. In this category we count first the number of abstract elements or types (\#AET), number of substitution groups (\#SG) and number of complex types derived by restriction or extension (\#TD). The values of these metrics are shown in Table \ref{tab7}.

\begin{table}[!h]
 \begin{center}
 \caption{Use of subtyping mechanisms}  \label{tab7}
   \begin{tabular}{| p{0.8cm} | p{0.8cm} | p{0.8cm} |p{0.8cm} |p{0.8cm}|p{0.8cm} |p{0.7cm} | }
     \hline
   \T \B   & SOS 1.0.0 & WFS 2.0 & WCS 2.0 & SPS 1.0.0 & WPS 1.0.0 & WMS 1.3.0 \\ \hline
     \#AET    & 61   & 15    & 74    & 52 &   2   & 2 \\ \hline
     \#SG     & 69   & 11   & 123   & 61   & 2   & 0 \\ \hline
     \#TD     & 349  & 56  & 371  & 297  & 31  & 4 \\ \hline
   \end{tabular}
\end{center}
 
 \end{table}

The results show that subtyping mechanisms are widely used in the specifications leading to an elevated number of non-explicit dependencies between schema components. This may lead to inadvertently overlook important details when analysing dependencies. 

\begin{sloppypar}
In order to measure in greater detail the influence of the subtyping mechanisms on complexity we introduce three new metrics: \textit{Data Polymorphism Rate(DPR)}, \textit{Data Polymorphism Factor} (DPF) and \textit{Schemas Reachability Rate} (SRR). 
\end{sloppypar}

The term \textit{Data Polymorphism (DP)} refers to the fact that nodes in XML instance files have a \textit{declared type}, but also have a \textit{dynamic type}. This is because global elements can be subtituted by any element in its substitution group. Similarly, an element in an instance file may be of any type derived from the declared type. This situation is similar to \textit{polymorphism} in the Object-Oriented Programming (OOP) context.

\subsubsection{Data Polymorphism Rate}

The \textit{Data Polymorphism Rate (DPR) }is a measure of how much polymorphism is contained in the schemas. It is expressed by the formula:

 \begin{equation} \label{eq3}
\begin{split}
DPR = \frac{\sum_{i=1}^{N}PE_{CTi}}{\sum_{j=1}^{N}E_{CTj}}
\end{split}
\end{equation}

In the formula, $N$ is the total number of complex types, $PE_{CTi}$ is the number of elements in the declaration of the complex type $CT_{i}$ that are polymorphic, i.e., its dynamic type may differ from its declared type in instance files. $E_{CTj}$ is the number of elements in the type declaration of type $CT_{j}$. For every type, a reference to a global element and an inner element declarations are considered as equals and count as 1. As a consequence, the numerator is the total number of polymorphic elements on the schemas.  Similarly,  the denominator is total number of elements contained in all complex types in the schemas. 

The result value is in the interval [0, 1], indicating the fraction of the elements that are polymorphic. This metric is a variation of the Polymorphic Factor (POF) metric used in the OOP context \cite{proc:abreu}.

\begin{table}[!h]
 \begin{center}
 \caption{DPR values for the OWS specifications} \label{tab8}
   \begin{tabular}{| p{1.5cm} | p{0.7cm} | p{0.7cm} |p{0.7cm} |p{0.7cm}|p{0.7cm} |p{0.7cm} | }
     \hline
   \T \B   & SOS 1.0.0 & WFS 2.0 & WCS 2.0 & SPS 1.0.0 & WPS 1.0.0 & WMS 1.3.0 \\ \hline
     DPR   & 0.13   & 0.12   & 0.15    & 0.13 &   0.05   & 0 \\ \hline
   \end{tabular}
\end{center}
 \end{table}

From the results shown in Table \ref{tab8}  we can observe that simpler specifications contains zero or a low degree of polymorphism. The rest of the specifications have a similar degree ranging between 12 and 15\%. Saying if these values are too high or not is not a trivial task, however in \cite{proc:abreu}, analysing polymorphism in the context of OOP is stated that a values of POF above 10\% is expected to reduce the benefits obtained with an appropriate use of polymorphism.  This is because highly polymorphical hierarchies will be harder to understand, debug and maintain.

\subsubsection{Data Polymorphism Factor}

The previous metric gives an idea of the number of polymorphic elements on the schemas, but does not measure their influence in the overall schemas complexity. For instance, a polymorphic element that can be substituted by two other elements does not have the same effect in complexity as an element that can be substituted by twenty different elements. In this regard, we define the \textit{Data Polymorphism Factor(DPF})as follows:

 \begin{equation} \label{eq4}
\begin{split}
DPF = \frac{\sum_{i=1}^{N}OE_{CTi}}{\sum_{j=1}^{N}E_{CTj}}
\end{split}
\end{equation}

In this case, $OE_{CTi}$ is the number of possible different elements that could be contained in a complex type. It is the summation of the number of elements declared in $CT_{i}$, the elements in the substitution groups of those elements, and the number of possible dynamic types that can have any element in $CT_{i}$ different from its declared type. For example, $OE_{ContainerType} = 3$ in Figure 1, because it contains two element declarations and one of them, \textit{containerElement}, may have a different dynamic type: \textit{ChildType}. The denominator is the same as in the definition of the DPF metric. In the formula $OE_{CTi} >= E_{CTi}$  for all i, natural number in the interval $[1,N]$. As a consequence the values of DPF are always equal or greater than 1, representing the factor in which the number of elements to be considered might grow when polymorphic elements are taken into account.

\begin{table}[!h]
 \begin{center}
 \caption{DPF values for the OWS specification schemas} \label{tab9}
   \begin{tabular}{| p{1.5cm} | p{0.7cm} | p{0.7cm} |p{0.7cm} |p{0.7cm}|p{0.7cm} |p{0.7cm} | }
     \hline
   \T \B   & SOS 1.0.0 & WFS 2.0 & WCS 2.0 & SPS 1.0.0 & WPS 1.0.0 & WMS 1.3.0 \\ \hline
    DPF    & 2.20   & 1.47   & 1.48    & 2.20 &   1.05   & 1 \\ \hline
   \end{tabular}
\end{center}
 \end{table}

Table \ref{tab9} shows the value of the metric for the schemas of the different specifications. These results show that the effect of polymorphic elements on SOS and SPS is higher that in WFS and WCS. Presumably this is caused by the larger number of dependencies of the sensor related specifications. SOS and SPS have a lot of common dependencies that is why DPF and DPFA values are basically the same for both specifications. As the simplest specifications barely contain polymorphical elements, the values of DPF for them are  equal or close to the minimal value, 1. 

\subsubsection{Schemas Reachability Rate}
The last metrics proposed in this paper, \textit{Schemas Reachability Rate (SRR}), attempts to measure the fraction of \textit{imported schema components} that are hidden (not referenced explicitly) by the subtyping mechanisms. As mentioned in Section 3, OWS specification schemas reutilise other specification schemas by \textit{importing} them in the main specification schemas. An imported component may be referenced directly if it is explicitly mentioned in the declaration of a schema component in the main schemas. But,  it can also be referenced \textit{indirectly}, if it is for example in the substitution group of a referenced element, or its derived from a type that is referenced directly. For example, it is not clear for everybody that if we are using GML 3.1.1 in our schemas and we define an inner element to be of type \textit{gml:AbstractFeatureType}, this element may have 13 different dynamic types (considering only GML types) on instance files based on these schemas.

To calculate SRR, we define first $G_{S}$, $G_{SH}$ and $V_{Rm}(G)$ as follows:\\

\textit{\textbf{Definition 1}: We define $G_{S}$  for the schemas in a specification $S$ as the directed graph $ G_{S} =(V_{S}, E_{S})$, where vertices in $V_{S}$, are all of the global elements declared in all of the schemas related to $S$ (main and external schemas). $E_{S}$ are directed edges between these vertices. An edge from $V_{i}$ to $V_{j}$ exists if $V_{j}$ is used somehow in the declaration of $V_{i}$.} \\

\textit{\textbf{Definition 2}: We define  $G_{SH}$  for the schemas in a specification $S$ as the directed graph $ G_{SH} =(V_{SH}, E_{SH})$, where $V_{SH} = V_{S}$. $E_{SH}$ extends $E_{S}$  by including also non-explicit dependencies, i.e. an edge from $V_{i}$ to $V_{j}$ exists if $V_{j}$ is used somehow in the declaration of $V_{i}$, or if $V_{j}$, is in the substitution group of an element referenced from $V_{i}$, or $V_{j}$ is a type derived from a type used in the declaration of $V_{i}$.} \\

\textit{\textbf{Definition 3}: We define, $V_{Rm}(G)$ for a directed graph $G = (V,E)$ and $V_{m}$, a subset of $V$, as the subset containing all of the vertices in $V$ that are reachable from at least a vertex in $V_{m}$} \\

Based on these definitions, if we consider that $V_{m}(G)$ is the subset of $V(G)$ containing the schema components included in the main schemas,  $V_{Rm}(G)$ would contain any schema component that is reachable from the main schemas. In the case of $G_{S}$, this will be components reachable through explicit dependencies, and in the case of $G_{SH}$, these are reacheable components through explicit and non-explicit dependencies. Using these vertex sets the SRR metric is calculated as follows:

\begin{equation} \label{eq5}
\begin{split}
SRR = \frac{|V_{Rm}(G_{SH})| - |V_{Rm}(G_{S})|}{ |V_{S}|}
\end{split}
\end{equation}

\begin{sloppypar}
The metric measures the fraction of schema components a specification depends from, but that are not explicitly referenced from any component in the main schemas, or any component reachable through explicit dependencies from the main schemas. 

\begin{figure}[!h]
  \centering
  \includegraphics[scale=0.095]{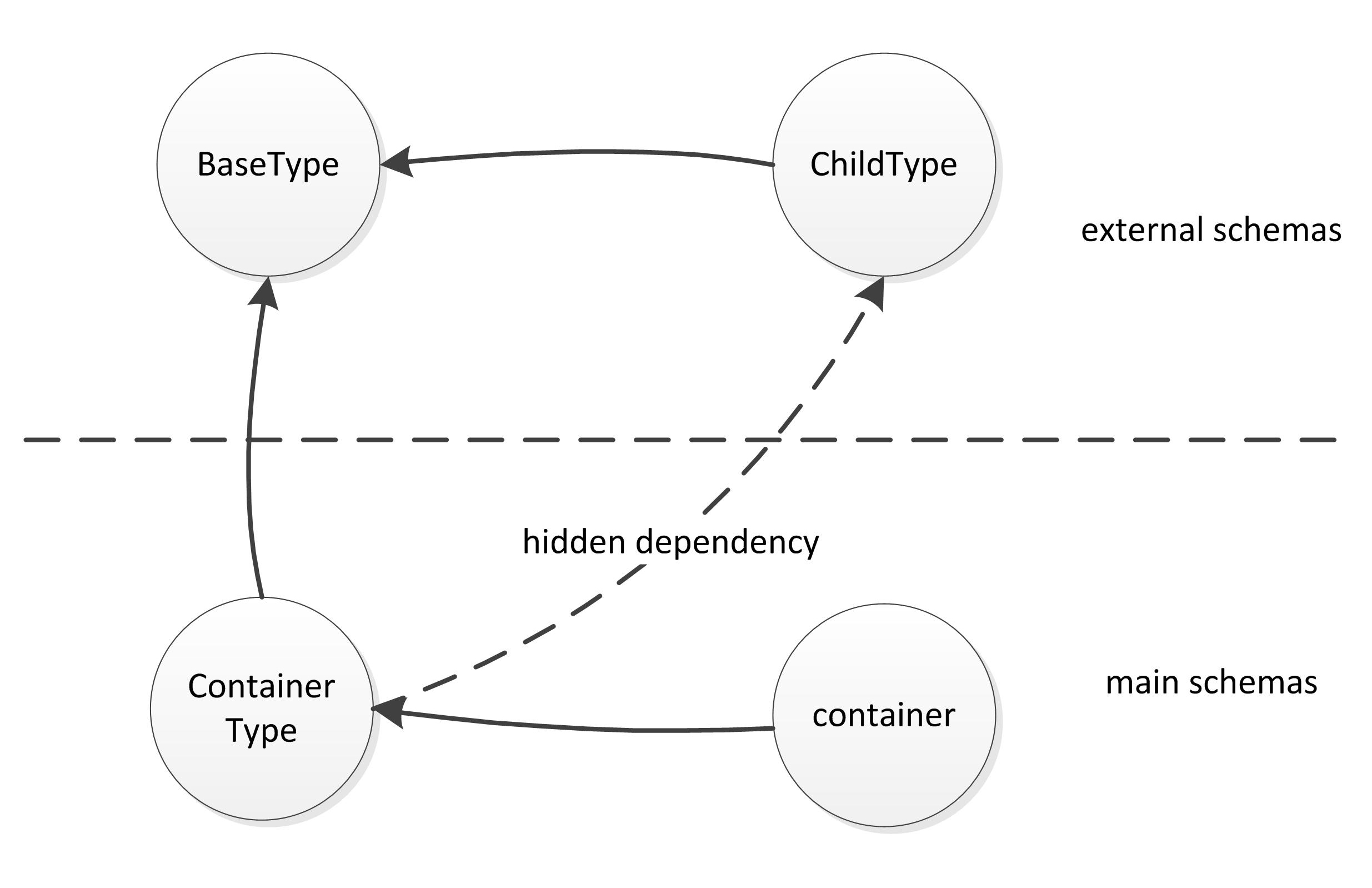}\\
  \caption{Graph of schema component relations for schema fragment in Figure 1}\label{fig5}
\end{figure} 

Figure \ref{fig5} shows the graph of component relations for the schema fragment in Figure \ref{fig1}. If we ignore the hidden dependency between \textit{ContainerType} and \textit{ChildType} we have $G_{S}$, otherwise we have $G_{SH}$. If we consider, for example, that the declaration of element \textit{container} and type \textit{ContainerType} are located in the main schemas, and  \textit{BaseType} and \textit{ChildType} declarations are located in external schemas we could calculate the value of DPRF for the main schemas: $V_{m} = \{container, ContainerType\}$, $V_{Rm}(G_{S}) = \{container, ContainerType, BaseType\}$, $V_{Rm}(G_{SH})  = \{container, ContainerType, BaseType, ChildType\}$, so:
\end{sloppypar}

\[
SRR = \frac{4 - 3}{4} = 0.25
\]

This value means that a quarter of the schema components are referenced through non-explicit dependencies.

Table \ref{tab10} shows the value of the cardinalities of the involved vertex sets and the value of the SRR metric for the schemas of the different specifications. The results shows that for SOS, WCS, and SPS more than 60\% of the schema components that could be used in instance files are not referenced explicitly from the schema component in the main schemas, or any component that is referenced from them. This high rate suggests that the effect of the subtyping mechanism in schemas complexity is enormous. For the rest of the specification the effect goes from moderate (WFS, WPS) to non-existent (WMS).

\begin{table}[!h]
 \begin{center}
 \caption{SRR values for the OWS specifacion schemas} \label{tab10}
   \begin{tabular}{| p{1.65cm} | p{0.7cm} | p{0.65cm} |p{0.7cm} |p{0.7cm}|p{0.65cm} |p{0.65cm} | }
     \hline 
                      \T \B     & SOS 1.0.0 & WFS 2.0 & WCS 2.0 & SPS 1.0.0 & WPS 1.0.0 & WMS 1.3.0 \\ \hline
     $|V_{Rm}(G_{SH})|$    & 1277  & 321       & 1349      & 1058       & 146  & 71 \\ \hline
    $|V_{Rm}(G_{S})|$     	 & 319    & 245    	 & 220        & 203         & 126   & 71 \\ \hline
    $|V_{S}|$   		         & 1498   & 353	 	 & 1625      & 1266       & 179   & 80 \\ \hline
     SRR     		             & 0.64   & 0.22	 	 & 0.69       & 0.68        & 0.11  & 0 \\ \hline
   \end{tabular}
\end{center}
 \end{table}

\section{Related Work}
Literature about measuring XML schemas complexity has increased in the last few years, based mainly on adapting metrics for assessing complexity on software systems or XML documents \cite{article:barbosa} \cite{article:mccabe} \cite{article:qureshi}. To our best knowledge the most relevant attempt in this topic is presented in \cite{proc:lammel}. Here, a comprehensive set of metrics is defined and applied to a large corpus of real-world XML schemas. Based on the resulting metrics, the authors define a categorization for a set of schema files according to its size. Another relevant study is \cite{proc:mcdowell} which defines eleven metrics to measure the quality and complexity of XML Schemas. 

In \cite{article:basci}, the authors present a more sophisticated metric that takes into account, not only the number of main schema components like the previous mentioned works, but also the internal structure of these components. In a similar way, \cite{proc:visser} proposes more advanced schema metrics, arguing that previous work on the topic only measure size as an approximation for complexity. The authors present a set of metrics to measure other structural properties of the schemas. 

Last about schemas complexity, in \cite{proc:pichler} the authors present a set of schema metrics in the context of schema mapping. A combined metric is defined based in simpler metrics considering schemas size, use of different schema features, and naming strategies. The combined metric is evaluated in the context of business document standards.

Similar studies in the geospatial domain are scarce, though, an interesting discussion of complexity can be found in \cite{blog:lake1} \cite{blog:lake2}. This discussion tries to identify the origin of GML complexity and use some of the metrics in \cite{proc:lammel} to categorize its schemas. Our research attempts to extend this discussion to the OWS specifications, but focusing more in the complexity of the schemas themselves. 

It is also worth mentioning that the problem of schemas complexity has been usually dealt with by using XML data binding code generators \cite{book:mclaughlin} or by using \textit{schema profiles} \cite{article:singh}. The first solution allows the automatic generation of XML processing code. Although these generators generally produce acceptable results, in presence of large schemas they may produce code that is excessively large or do not meet performance application requirements. The second solution is based on extracting subsets of the schemas that are relevant to an actual implementation or problem domain. By using only a portion of the schemas, the complexity of handling or understanding them is reduced in a large degree. Examples of profiles in the context of GML are the Simple Feature Profile\cite{ogc:gmlfeatureprofile} and the Common CRS Profile \cite{ogc:gmlcommoncrsprofile}.

\section{Conclusions}
In this paper we have presented a quantitative way to analyse and measure the complexity of OWS' schemas. The results of the analysis have shown that at least half of the presented specifications can be considered as large and complex according to all of the metrics included in our study. Most of the metrics coincide in finding a clear differentiation between a first group containing large specifications (SOS, SPS and WCS), a second group containing medium sized specifications (WPS and WFS), and a third group of simple specifications (WMS).   More complex specifications, as a general rule, are those that present a larger number of dependencies from other specifications.
 
The new metrics introduced here have shown from different views  the effect of the use of subtyping mechanism on complexity. For example, DPR has shown the fraction of polymorphic elements, frequently high, included in the schemas. DPF has considered how the possible polymorphic situations for these elements increase the effort needed to fully understand the schema components definitions. Last, SRR has shown that more than 60\% of the schema components included in large specifications are referenced in ways that cannot be seen explicitly, augmenting the risk of making mistakes while working with the schemas.

The metric set presented here should not be seen as a closed set, many other metrics can be useful in many different scenarios. The use of adequate metrics allows us to quantify the complexity, quality and other properties of the schemas. This can be very useful in different scenarios, such as, evaluating the impact of design decisions, assessing the effectiveness of different solutions to deal with  schemas complexity (e.g. how the use of schema profiles simplifies the implementation and comprehension of a given problem). Metrics can also be very useful to detect potential design problems such as components with to many information items, or excessively deep subtyping hierarchies, just to mention some examples. Last, metrics can also suggests solutions about how to deal with the large size and complexity of geospatial schemas.

\section{Acknowledgements}
This work has been partially supported by the ``Espa\~{n}a Virtual'' project (ref. CENIT 2008-1030) through the Instituto Geogr\'{a}fico Nacional (IGN). 

%

\bibliographystyle{abbrv}
\bibliography{article}  

\balancecolumns

\end{document}